# Quantum compressed sensing tomographic reconstruction algorithm

Arim Ryou, Kiwoong Kim, and Kyungtaek Jun

A. R. is with Department of Physics, Chungbuk National University, Chungcheongbuk-do, 28535, South Korea (e-mail: arimryou01@gmail.com, arimryou@chungbuk.ac.kr).

K. K. is with Department of Physics, Chungbuk National University, Chungcheongbuk-do, 28535, South Korea (e-mail: kiwoong@chungbuk.ac.kr).

K. J. is with Quantum Research Center, QTomo, Chungcheongbuk-do, 28535, South Korea and Chungbuk Quantum Research Center, Chungbuk National University, Chungcheongbuk-do, 28644, South Korea (e-mail: ktfriends@gmail.com, ktfriends@chungbuk.ac.kr).

(Corresponding author: Kiwoong Kim and Kyungtaek Jun.)

## Abstract

Computed tomography (CT) is a non-destructive technique for observing internal images and has proven highly valuable in medical diagnostics. Recent advances in quantum computing have begun to influence tomographic reconstruction techniques. The quantum tomographic reconstruction algorithm is less affected by artifacts or noise than classical algorithms by using the square function of the difference between pixels obtained by projecting CT images in quantum superposition states and pixels obtained from experimental data. In particular, by using quantum linear systems, a fast quadratic unconstrained binary optimization (QUBO) model formulation for quantum tomographic reconstruction is possible. In this paper, we formulate the QUBO model for quantum compressed sensing tomographic reconstruction, which is a linear combination of the QUBO model for quantum tomographic reconstruction and the QUBO model for total variation in quantum superposition-state CT images. In our experiments, we used sinograms obtained by using the Radon transform of Shepp-Logan images and body CT images. We evaluate the performance of the new algorithm by reconstructing CT images using a hybrid solver with the QUBO model computed from each sinogram. The new algorithm was able to obtain a solution within 5 projection images for $30 \times 30$ image samples and within 6 projection images for $60 \times 60$ image samples, reconstructing error-free CT images. We anticipate that quantum compressed sensing tomographic reconstruction algorithms could significantly reduce the total radiation dose when quantum computing performance advances.

**Index Terms** – quantum compressed sensing, quantum tomography, quantum optimization, quantum tomographic reconstruction, quantum compressed sensing tomographic reconstruction, QCSTR algorithm

## Introduction

Quantum algorithms using quantum bits (qubits) have evolved along with the development of quantum computers/annealers. Quantum algorithms can perform more efficient calculations than classical algorithms because they can be exponentially faster with the number of qubits used by using the superposition and/or entanglement of qubits [1,2,3]. Representative examples used in gate-based quantum computers include Shor's algorithm [4], which can exponentially speed up factoring, Grover's algorithm [5], which improves search speed by two orders of magnitude compared to existing algorithms, and the HHL algorithm [6], which can predict the features of solutions to linear systems. The quantum annealing-based D-Wave system has been able to compute 2 million variables for a quadratic unconstrained binary optimization (QUBO) model by developing a hybrid solver [7,8,9]. As the performance of quantum annealers capable of computing QUBO models increases, there have been approaches to formalizing QUBO models for various NP-hard problems [10,11,12,13]. Additionally, quantum machine learning, which generalizes certain patterns that cannot be generalized on classical computers but can be generalized on quantum computing, is one of the most promising fields [14,15,16,17].

Advances in the D-Wave system's hybrid solver nearly find the global minimum energy of a QUBO model using 10,000 variables. These developments have made it possible to efficiently compute QUBO models. Recently, the QUBO model for quantum linear systems, one of the most fundamental problems in modern engineering and science, has been proposed [18]. In linear system computation, classical algorithms require $O(n^3)$ computational cost, but quantum algorithms can compute QUBO formulations derived from a linear system in at most $O(\log_2 n)$ through parallel computation. Quantum linear systems enabled us to compute the QUBO model for computed tomography (CT) image reconstruction [19]. The quantum tomographic reconstruction (QTR) algorithm can be used in spiral computed tomography (CT) [20], electron tomography [21], and synchrotron X-ray tomography [22]. In particular, it can be used for cone-beam, fan-beam, and parallel-beam light sources by calculating the projected geometry of the tomographic system. For each system, various classical tomographic reconstruction algorithms have been developed [23,24,25]. In experiments using image samples, classical algorithms were able to reconstruct blurry CT images, but QTR algorithm showed the potential to reconstruct clean CT images [26]. By using the difference of X-ray mass attenuation coefficients (MACs) [27] instead of the radix 2 representation [28] when representing pixels in a quantum superposition state, the hybrid solver succeeds in error-free quantum CT images by finding the global minimum energy [29]. QTR algorithm has severe limitations in the number of qubits that can be used, and quantum algorithms are being developed to decompose the qubits used to solve this problem [30,31]. Despite these developments, the total radiation dose received from CT scanning is still considered a concern.

In this paper, we propose a novel QUBO model that combines a QTR algorithm [19] and a quantum compressed sensing algorithm. We use the hybrid solver for Constrained Quadratic Models (CQM) and the Binary Quadratic Models (BQM) to combine the two QUBO models. Each optimization operation used in the quantum tomographic reconstruction algorithm can be expressed as a first-order operation in the CQM solver and as a second-order operation in the BQM solver. A newly proposed quantum compressed sensing tomographic reconstruction (QCSTR) algorithm fails to satisfy constraints in CQM solver calculations, but achieves good results in BQM solver. To obtain the experimental results, we used image samples with X-ray MACs of 1, 1 and 2, and 1, 2, 3. The size of the image sample was set to $30 \times 30$ and $60 \times 60$, respectively, with each X-ray MAC. The sinogram obtained from the CT system can be subjected to similar conditions through image preprocessing [29]. We were able to reconstruct CT images with no or small errors using between 10% and 20% of the number of projection images that would normally be required. The new algorithm shows very important results because it can reconstruct CT images that are similar to real structures while reducing the total radiation dose during CT scanning. In particular, the QTR algorithm can be used in all tomographic systems and shows useful results even on data with noise or artifacts [30,32]. The QCSTR algorithm includes all the advantages of the QTR algorithm. We believe that quantum CT images will play a major role in the advancement of medical imaging diagnostics and medical automation because of their quantum supremacy.

## Method

### Background for a quantum tomographic reconstruction algorithm

The QUBO model is computable on gated model quantum computers and quantum annealers. The QUBO model can obtain the optimal energy in the gate model quantum computer and the quantum annealer. The QUBO model used in the calculation using QAOA [33] and the QUBO model calculated in the quantum annealer are only different in sign [19]. The hybrid solver in the D-Wave system is capable of computing global minimum energy for 10,000 by 10,000 QUBO matrices [29], which is sufficient to demonstrate the results of the QUBO model on test samples. Therefore, we focus on the QUBO model for the global minimum energy computable by a hybrid solver.

The pixel values measured within charge-coupled devices (CCD), typically in CT systems, decrease as X-rays pass through the sample and are related to the accumulation of the X-ray MACs of the sample. In this paper, to simplify the problem, we proceed under the condition that an ideal sinogram can be obtained from a CCD. Let $P(\theta, s)$ be the measurement for projection angle $\theta$ and position $s$ within the CCD. In order to visualize the QUBO model, the line integral with respect to $P(\theta, s)$ must be performed according to the type of X-ray source. The light sources of the parallel-beam and fan-beam can be calculated in the form of an integral for 2D, but the cone-beam requires an integral for 3D. We formulate the QUBO model for 2D to simplify the problem, but it is possible to compute the QUBO model for 3D in a similar way.

Let the pixel value at position $(i, j)$ of the quantum superposition state CT image be $I_{ij}$. If the pixels affected by $P(\theta, s)$ are $I'_{ij}$, the density $IP(\theta, s)$ of the quantum superposition state according to the projected

light source can be calculated as follows:

$$IP(\theta, s) = \sum_{i,j} c_{ij} I'_{ij} \tag{1}$$

where $c_{ij}$ represents the overlapping area according to the light source. The energy function $F$ for tomographic reconstruction can be expressed as follows:

$$F_1 = \sum_{\theta=0}^{180-d\theta} \sum_{s=1}^{n} ((IP-P)(\theta,s))^2 - \sum_{\theta=0}^{180-d\theta} \sum_{s=1}^{n} (P(\theta,s))^2 \tag{2}$$

The global minimum energy for reconstructing error-free CT images is $-\sum_{\theta=0}^{180-d\theta} \sum_{s=1}^{n} (P(\theta,s))^2$.

To represent $I_{ij}$ in a quantum superposition state, there are three possible representations:

$$I_{ij} = \sum_{k=-m_1}^{m_2} 2^k q_k^{ij} \tag{3}$$

$$= \sum_{k=1}^{m} \alpha_k q_k^{ij} \tag{4}$$

$$= \alpha_1 q_1^{ij} + \sum_{k=2}^{m} (\alpha_k - \alpha_{k-1}) q_k^{ij} \tag{5}$$

Equation 3 uses the radix 2 representation [28], and Eqs. 4 and 5 use the X-ray MAC representation [29]. QUBO model $Q_1$ can be formulated by substituting one of Eqs. 3, 4, or 5 into Eq. 2.

## Quantum compressed sensing algorithm

Compressed sensing is a technique for achieving high-quality image reconstruction with less data, which is achieved by minimizing the total variation (TV) of the image, which is a measure of the sum of the gradient magnitudes [34]. In general, compressed sensing algorithms perform best in $L^1$-norm [35]. To obtain the QUBO model for energy minimization, we use the square function of the difference between two values. So, the QUBO model is expressed in the form of the $L^2$-norm squared. We use $I_{ij}$ in the form of Equation 5 to reduce the weakness of the QUBO model for compressed sensing in tomographic reconstruction.

Now, we can formulate the QUBO model for the TV of the image. The difference value between two adjacent pixels in a CT image in a quantum superposition state can be formulated as part of the QUBO model as follows:

$$\left(I_{i,j} - I_{i,j+1}\right)^2 = I_{i,j}^2 - 2I_{i,j}I_{i,j+1} + I_{i,j+1}^2 \tag{6}$$

To simplify the calculation, let $\alpha_k - \alpha_{k-1}$ be $\beta_k$ for $k \geq 2$ and let $\alpha_1 = \beta_1$. In Eq. 6, each square term is expressed as a sum of the linear and quadratic terms, as follows:

$$\left(\sum_k \beta_k q_k^{ij}\right)^2 = \sum_k \beta_k^2 q_k^{ij} + 2 \sum_{k_1 < k_2} \beta_{k_1} \beta_{k_2} q_{k_1}^{ij} q_{k_2}^{ij} \tag{7}$$

$I_{i,j}I_{i,j+1}$ is simply calculated as a sum of the quadratic terms.

The energy function $F$ of quantum compressed sensing for tomographic reconstruction is computed as the sum of the variations of all adjacent pixels, which can be expressed as follows:

$$F_2 = \sum_{i=1}^{n}\sum_{j=1}^{n-1}(I_{i,j} - I_{i,j+1})^2 + \sum_{j=1}^{n}\sum_{i=1}^{n-1}(I_{i,j} - I_{i+1,j})^2 \quad (8)$$

QUBO model $Q_2$ can be formulated by substituting one of Eqs. 3, 4, or 5 into Eq. 8.

## Quantum compressed sensing tomographic reconstruction algorithm

In tomographic reconstruction, the cubic model $Q$ using quantum compressed sensing can be expressed in a linear combination form as follows:

$$Q = aQ_1 + bQ_2 \quad (9)$$

where $a$ and $b$ are real numbers. The calculation for $Q_1$ can also use the CQM solver of the hybrid solver. When using the Discrete Quadratic Models (DQM) solver of the hybrid solver, it can be calculated using the energy function $F = aF_1 + bF_2$. In this paper, we formulate the optimal QUBO model for the case where a and b are natural numbers to simplifying experiments and focus on computing the BQM solver of hybrid solver for $Q$.

## Result and implementation

This study was conducted on the number of projection images required for a QCSTR algorithm to reconstruct an ideal CT image from each ideal sinogram. Projection images were obtained at isometric angles from 0 to $180 - d\theta$ degrees, where $d\theta$ represents the change in angle. To test the new algorithm, we used the QUBO model of Eq. 9 after quantum superposition of each pixel in the CT image using Eq. 5 for the X-ray MAC representation. Each experiment used the BQM solver of the hybrid solver of the D-Wave system. To simplify the presentation of our results, we omit the sinograms and focus on image samples and CT images reconstructed with the new algorithm. After we found that the QUBO model with $a$ and $b$ fixed obtained the global minimum energy we wanted, we ran four more experiments to verify that the new algorithm found the same energy. Each test image sample was a high-quality image resized to a smaller size and then Gaussian blurring was used. We assumed that the X-ray MAC of the samples used in each experiment was already known.

### A. Shepp-Logan phantom image

The sample used in the first experiment was a $30 \times 30$ Shepp-Logan phantom image with one X-ray MAC (see Fig. 1a). This image was generated from the scikit-image library and then resized. For a single material sample, the X-ray MAC of the sample was set to 1 because the MAC can always be made 1 by dividing $\alpha$ in the sinogram. In this experiment, three projections were used in the QUBO model formulation. The global minimum energy of the QUBO model desired by the new algorithm is $-14460.70$, and when $(a, b) = (1, 1)$, we obtain $-14460.70$. The global minimum energy of the QUBO model desired by the new algorithm is $-14460.70$, and we obtain exactly the same energy to two decimal places when $(a, b) = (1, 1)$. At this time, the TV is 150 and the reconstructed CT image is error-free (see Fig. 1a). As b increased, more monotonous CT images were reconstructed (see Figs 1b and c). The QUBO model using only $Q_1$ for QTR algorithm is shown in Fig. 1d.

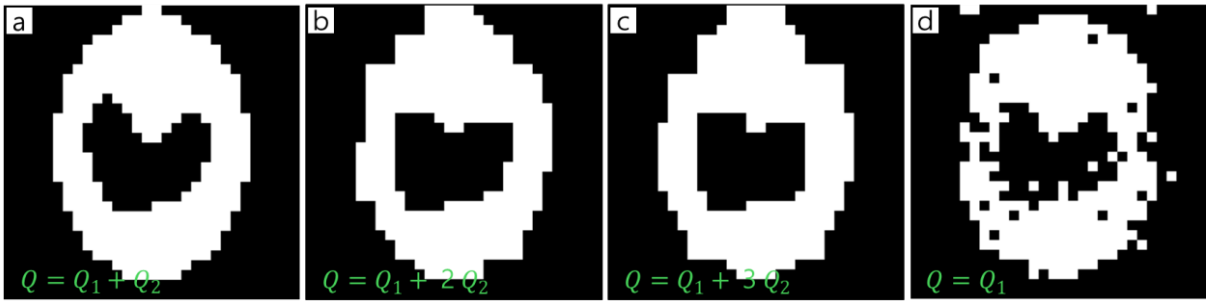

Figure 1. **30 × 30 CT images for a Shepp-Logan image sample with one X-ray MAC**. In this experiment, three projection images with projection angles varying by 60 degrees were used. (a) An image sample and a CT image reconstructed by the QCSTR algorithm for the QUBO model, (b, c) CT images reconstructed by the QCSTR algorithm for each QUBO model, (d) A CT image reconstructed by the QTR algorithm.

The sample used in the second experiment was a 30×30 Shepp-Logan phantom image containing three X-ray MACs: 1, 2, and 3 (see top right panel in Fig. 2). We conducted experiments on QUBO model $Q = Q_1 + Q_2$ and increased the number of projection images from 3 to 5. The global minimum energy of the QUBO model desired by the new algorithm is $-166962.22$, we obtain exactly the same energy when five projection images were used. At this time, the TV is $302$ and the reconstructed CT image is error-free. Figure 2 shows the reconstructed images according to the number of projection images by the QCSTR algorithm and the QTR algorithm without the compressed sensing constraint.

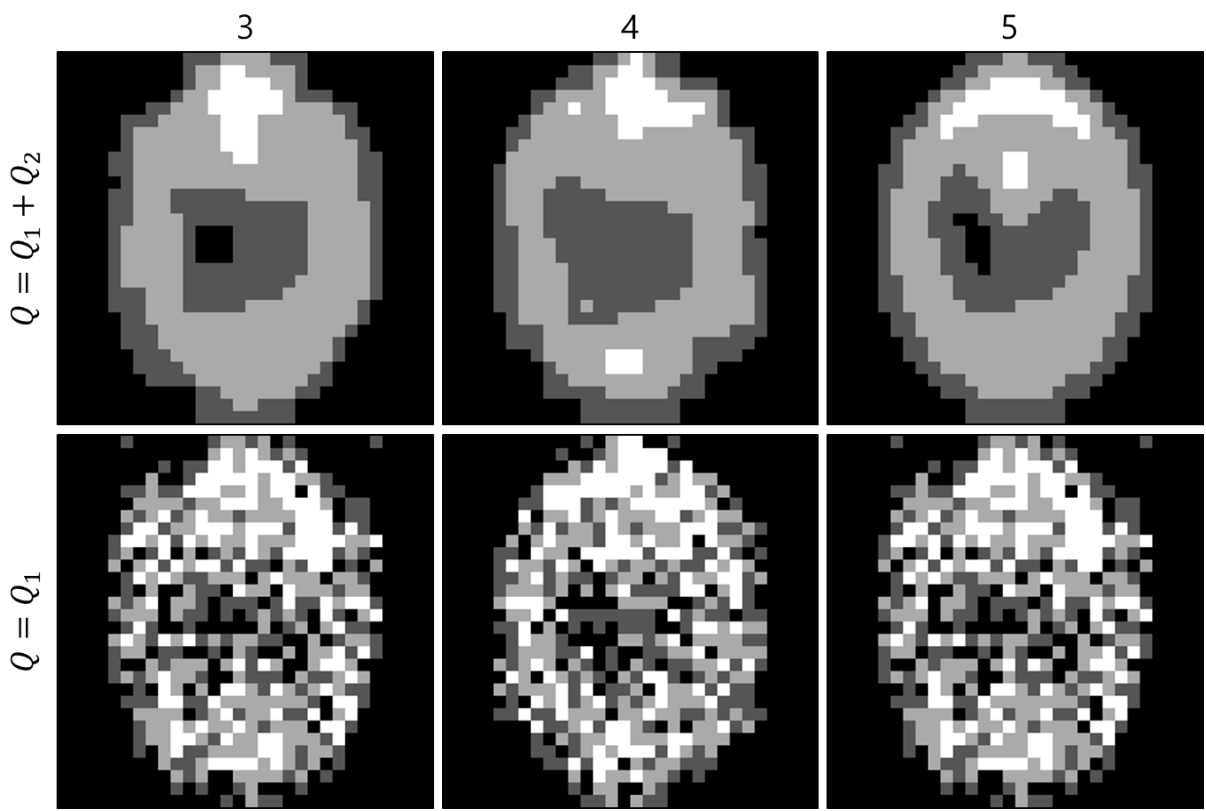

Figure 2. **30 × 30 CT images for a Shepp-Logan image sample with three X-ray MACs: 1, 2, and 3**. In this experiment, the number of projection images was increased from 3 to 5. The upper row used the QCSTR algorithm for tomographic reconstruction, and the lower row used the QTR algorithm.

The following experiment was performed on a 60×60 Shepp-Logan phantom image with three X-ray MACs: 1, 2, and 3 (See 2nd panel from the left in Fig. 3). Six projection images were used in the QUBO model formulation. We varied $b$ to obtain optimal CT images. The global minimum energy of the QUBO model desired by the new algorithm is $-675518.50$, and when $(a, b) = (1, 2)$, we obtain $-675518.50$. At this time, the TV is 786 and the reconstructed CT image is error-free. Figure 3 shows the reconstructed images according to the number of projection images by the QCSTR algorithm and the QTR algorithm.

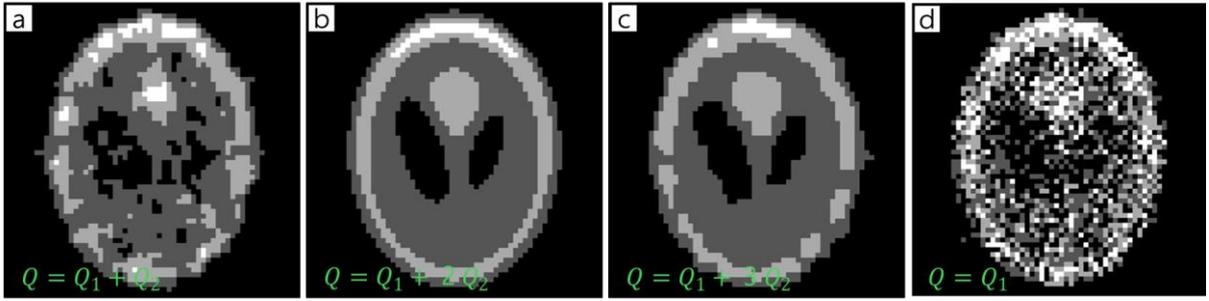

Figure 3. **60 × 60 CT images for a Shepp-Logan image sample with three X-ray MAC: 1, 2, and 3**. In this experiment, six projection images were used in the QUBO model formulation. (a, c) CT images reconstructed by the QCSTR algorithm for the QUBO model, (b) An image sample and a CT image reconstructed by the QCSTR algorithm for the QUBO model, (d) A CT image reconstructed by the QTR algorithm.

### B. Body CT image from Kaggle

A body CT image in Fig. 4a from the SIIM Medical Images dataset [36] was used in this experiment for analysis. We resized the body image to $30 \times 30$ and $60 \times 60$ images, as shown in Figure 4b and c, respectively, to ensure that the hybrid solver uses the number of qubits that can obtain the global minimum energy.

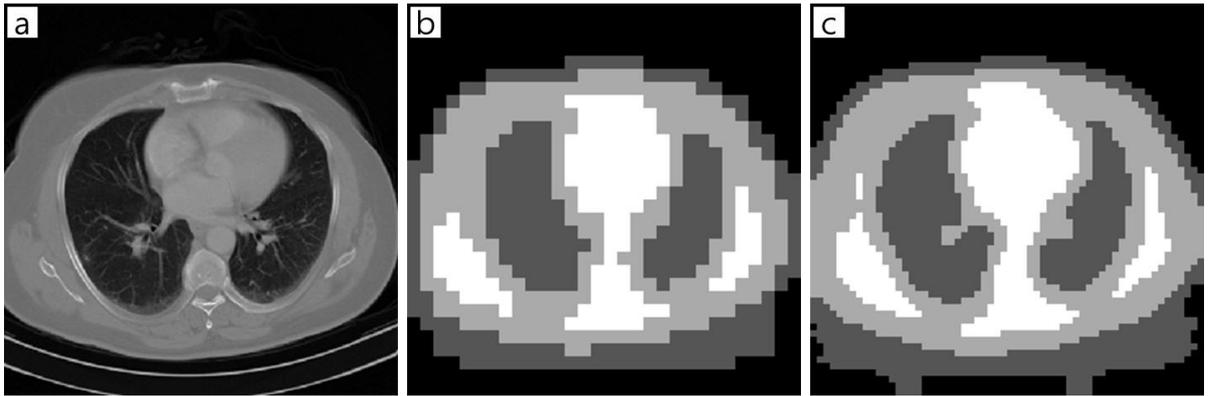

Figure 4. **A body CT image and image samples resized to $30 \times 30$ and $60 \times 60$.**

This experiment was conducted on a $30 \times 30$ body CT image containing three X-ray MACs: 1, 2, and 3 (see Fig. 4b). We conducted experiments on QUBO model $Q = Q_1 + Q_2$ and increased the number of projection images from 3 to 5. The global minimum energy of the QUBO model desired by the new algorithm is $-257589.71$, we obtain exactly the same energy when five projection images were used. At this time, the TV is 386 and the reconstructed CT image is error-free. Figure 5 shows the reconstructed images according to the number of projection images by the QCSTR algorithm and the QTR algorithm.

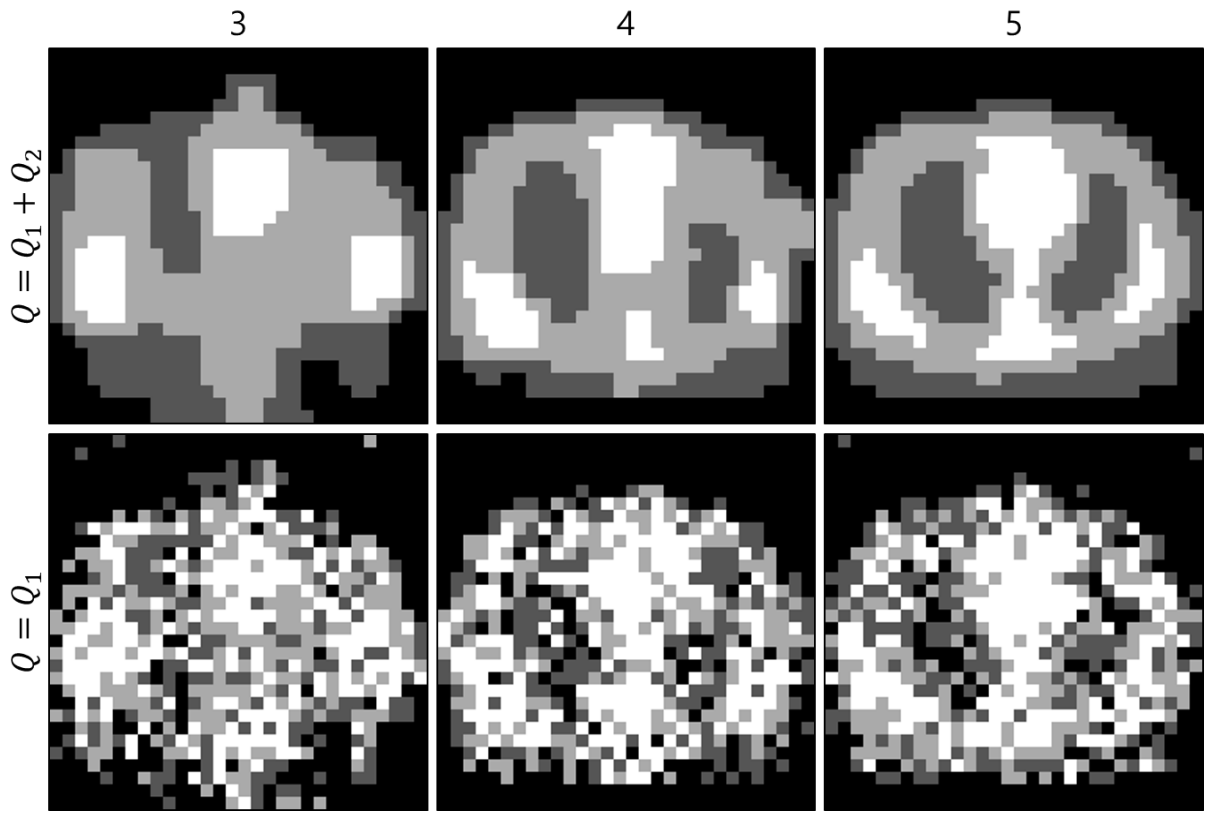

Figure 5. **30 × 30 CT images for a body image sample with three X-ray MACs: 1, 2, and 3**. In this experiment, the number of projection images was increased from 3 to 5. The upper row used the QCSTR algorithm for tomographic reconstruction, and the lower row used the QTR algorithm.

The next experiment was performed on a 60 × 60 body CT image with three X-ray MACs: 1, 2, and 3 (See Fig. 4c). Six projection images were used in the QUBO model formulation. We varied $b$ to obtain optimal CT images. The global minimum energy of the QUBO model desired by the new algorithm is $-2765350.00$, and when $(a, b) = (1, 2)$, we obtain exactly the same evergy. At this time, the TV is 835 and the reconstructed CT image is error-free. In the case of $(a, b) = (1, 3)$, there was an error of one pixel. Figure 6 shows the reconstructed images according to the number of projection images by the QCSTR algorithm and the QTR algorithm.

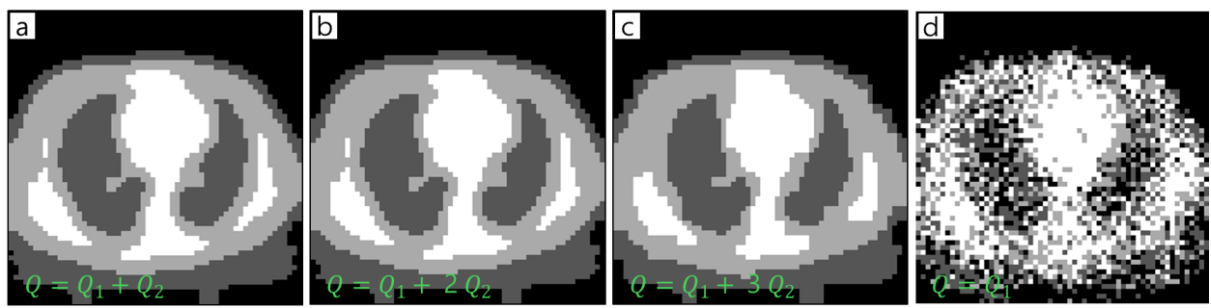

Figure 6. **60 × 60 CT images for a body CT image sample with three X-ray MAC: 1, 2, and 3**. In this experiment, six projection images were used in the QUBO model formulation. (a,c) CT images reconstructed by the QCSTR algorithm for the QUBO model, (b) An image sample and a CT image reconstructed by the QCSTR algorithm for the QUBO model, (d) A CT image reconstructed by the QTR algorithm

## C. Comparison of classical and quantum CT images

We compare CT images reconstructed with the QCSTR algorithm and previous quantum and classical algorithms. The test image sample used was the 60×60 Shepp-Logan phantom image in Fig. 3b. The sinogram obtained using the Radon transform on this sample is the data used in the experiment. This sinogram requires about 60 columns to satisfy the Helgason-Ludwing consistency condition. We used a sinogram consisting of six projection images, corresponding to about 10%. The classical algorithms, simultaneous algebraic reconstruction technique (SART) and filtered back projection (FBP), reconstructed the CT images shown in Figs. 7a and b. The reconstructed images using SART were obtained after 6 iterations, and the FBP-reconstructed images used the ramp filter. A CT image reconstructed with an early version of the QTR algorithm [17] is shown in Figure 7c. The global minimum energy of the QUBO model desired is −677090.50, the algorithm obtain exactly the same energy. The CT image reconstructed with the new algorithm is error-free and is shown in Fig. 7d. The new algorithm obtain −675518.50 when $(a, b) = (1, 2)$. At this time, the TV is 786 and the reconstructed CT image is error-free.

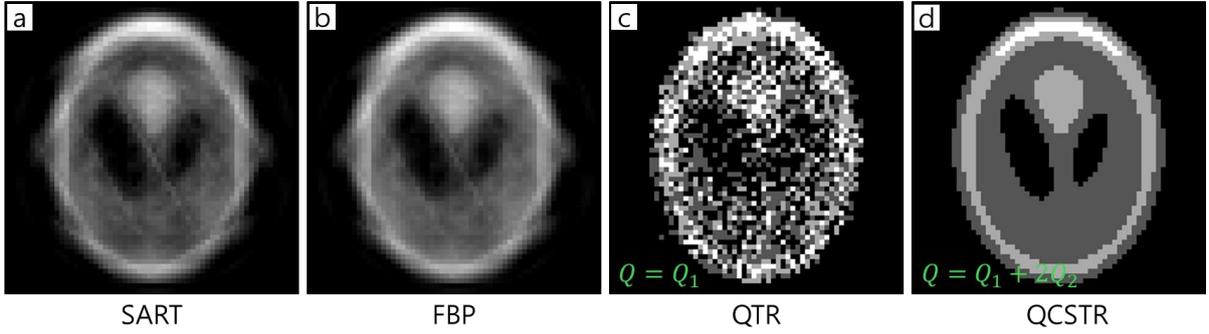

Figure 7. **Comparison of CT images.** A 60×60 Shepp-Logan phantom image was used as a test sample, and the sinogram was consisted of 6 projection images.

We have investigated the performance of a QCSTR algorithm on a real-world, noisy, and similar projection dataset. The noise generated from the CCD was assumed to have a random noise with a Gaussian distribution as shown in Eq. 10. [37].

$$I_i^* = I_i + \sigma_i z_i \qquad (9)$$

where $I_i^*$ is the x-ray intensity generated in the $i$th detector of a virtual CT, $\sigma_i$ is standard deviation, $I_i$ is the X-ray intensity, and $z_i$ is a standard normal random variable. The Poisson distribution is a common noisy model, but it is approximated by a Gaussian distribution with standard deviation $\sigma_i \sim I_0 \sqrt{I_i}$. We made the test sinogram data similar to the noisy sinogram measured at the synchrotron (See Supplementary Fig. 1) [29]. Each pixel in the noisy sinograms contained a random error of about 5%. Figure 8 shows CT images reconstructed by each algorithm. The first and second rows used the image in the upper right panel of Fig. 2 as a test sample, and the third and fourth rows used Fig. 3b. The absolute error between each CT image and the test sample is shown in Table 1. We additionally show the differences for each CT image for Fig. 8 in Supplementary Fig. 2.

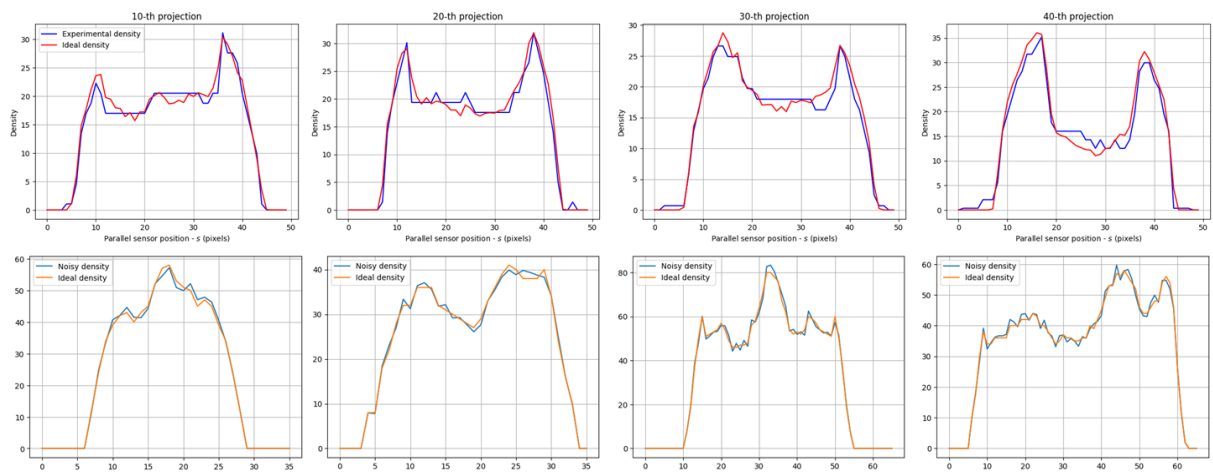

Supplementary Figure 1. **Density graph**. The upper images are data from the synchrotron radiation accelerator, and the lower images are data to which random noise with a Gaussian distribution was applied for this experiment.

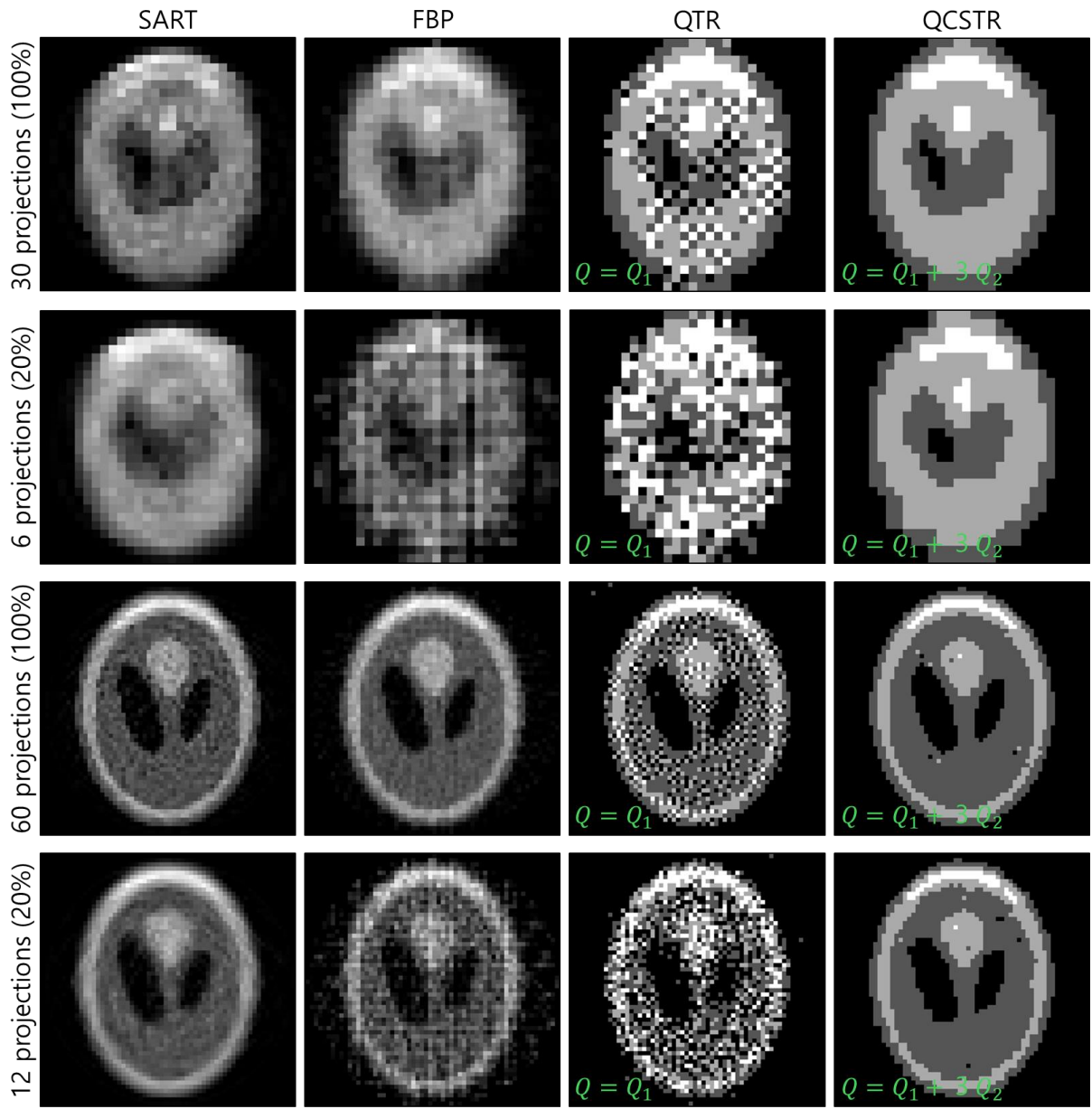

Figure 8. **Comparison of CT images reconstructed from noisy sinograms.** The upper two rows of CT images were reconstructed using the sinogram of the sample used in Fig. 2, and the lower two rows were reconstructed using the sinogram used in Fig. 3. In this experiment, 5% random noise with a Gaussian distribution was added to the ideal sinogram as shown in Supplementary Fig. 2.

|  | SART | FBP | QTR | QCSTR |
|---|---|---|---|---|
| 30 projections (100%) | 147.31 | 116.05 | 196.00 | 7.00 |
| 6 projections (20%) | 168.76 | 283.36 | 367.00 | 44.00 |
| 60 projections (100%) | 338.28 | 335.70 | 729.00 | 19.00 |
| 12 projections (20%) | 391.31 | 724.04 | 1218.00 | 103.00 |

Table 1. Absolute error of each CT image in Figure 8

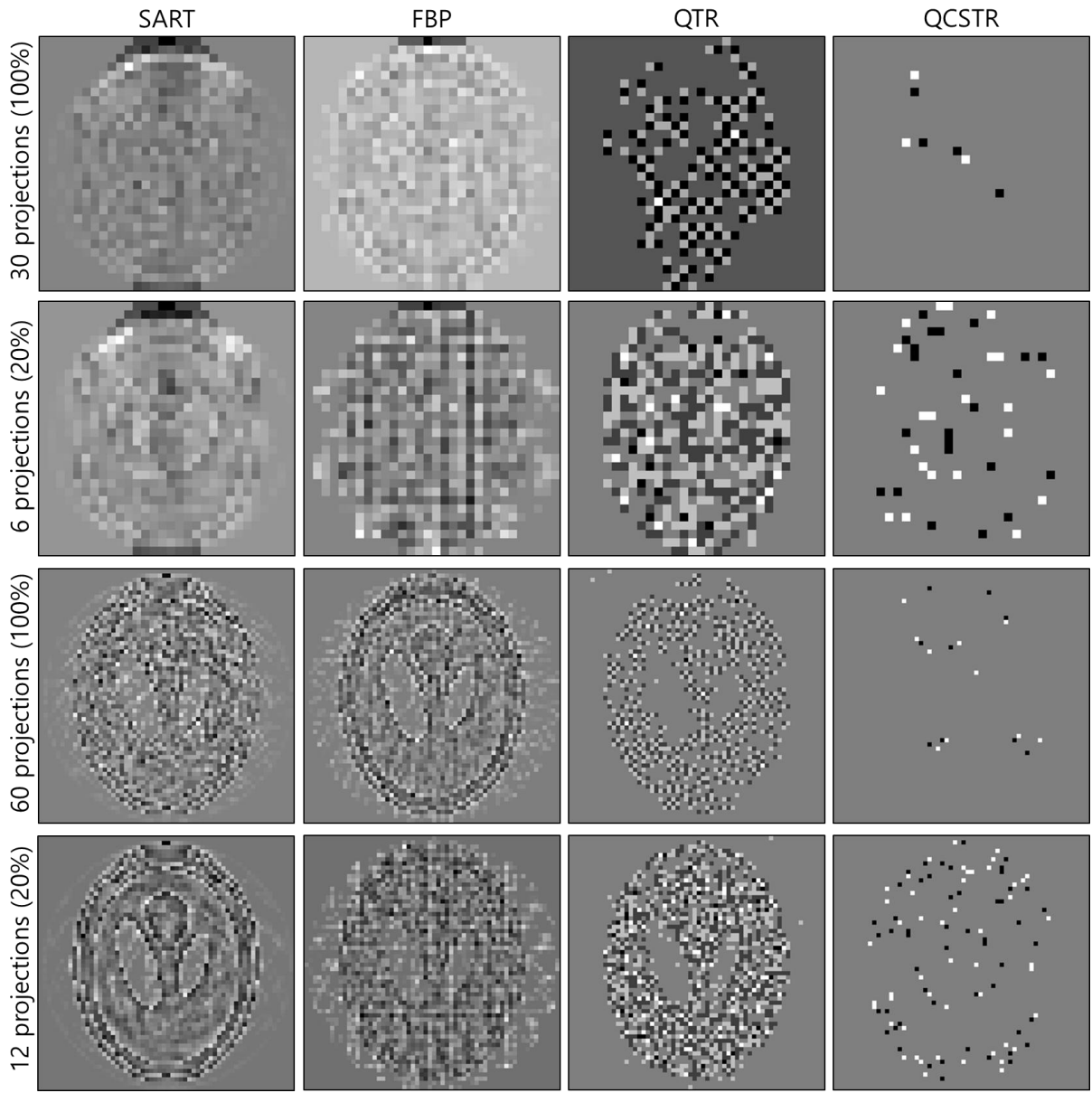

Supplementary Figure 2. **Difference between each sample and CT image in Figure 8.**

## Discussion

The hybrid solver has three solvers for BQM, DQM, and CQM. The QUBO model for QCSTR algorithm is computable by all three solvers. This algorithm uses two QUBO models for QTR and compressed sensing algorithms. The DQM solver can compute quadratic models even if they are not QUBO models. In this paper, BQM and CQM solvers were used because the QUBO model for tomographic reconstruction was used. The reason is that as quantum computers/annealers advance, we expect QPU solvers to provide more than 1 million logical qubits, and these solvers can only compute the QUBO model. CQM solver can use linear equations instead of quadratic equations for $IP(\theta, s) - P(\theta, s)$. Using a CQM solver, the calculations are slow, but it allows for precise control over each term, including errors. Additionally, since we do not square each optimization term in the process of formulating the QUBO model, we do not have to worry about the problem of the order increasing. However, since the QUBO models for QTR and quantum compressed sensing algorithms are completely different during the annealing process, satisfactory results could not be obtained when the QUBO model for QTR algorithm was used as a constraint. However, the global minimum energy shape satisfied by the QUBO model for QTR algorithm during the annealing process is completely different from that of the QUBO model for quantum compressed sensing, so when the QUBO model for QTR algorithm was used as a constraint, it was not possible to obtain results that satisfied all constraints. So, we used the BQM solver to linearly combine the two QUBO models.

In the QUBO model formulation for quantum compressed sensing, we use the whole square of the difference between two adjacent pixels. This means the square of the $L^2$-norm. It is generally known that compressed sensing algorithms are more suitable for $L^1$-norm [35]. To eliminate these weaknesses, we used the X-ray mac representation instead of the radix 2 representation. We used Eq. 5 and set up the experiment so that the L1 and L2 norms are equal while giving the condition that the X-ray MACs change by 1. Additionally, we did gaussian blurring on the resized samples used in the experiment to create image samples that changed as continuously as possible. And we set the image to have a size of $60 \times 60$ and 3 X-ray MACs so that the hybrid solver can use only up to 10800 variables. The new algorithm required up to five projection images for an $30 \times 30$ image to reconstruct an error-free CT image. In contrast, the new algorithm required about 10% of the projection image for a 60×60 image. We additionally experimented with $50 \times 50$ images and were able to achieve error-free tomographic reconstruction with approximately 10% of the projection images. We believe that if the X-ray MAC inside the sample varies continuously and we use CT images of a certain size or larger, the new algorithm will reconstruct an optimal CT image using about 10% of the projection images. In additional experiments, most of the results could be reconstructed error-free CT images when b was varied, but there were cases where results were obtained when a was varied. Of course, if the internal structure of the sample has many discontinuous changes or many fast-changing continuous pixels, more projection images are required. We are currently conducting research to relax the conditions on the internal structure of these samples.

As shown in Fig. 8, CT images reconstructed by SART, FBP, QCTR, and QCSTR algorithms were sequentially compared. The SART algorithm's error decreased as the number of iterations increased, but then started to increase again. In this experiment, the SART algorithm generally produced clean CT images when it was repeated 6 times, so this image was selected. So, when using X-ray projection data, errors are likely to be higher. We used the highest performing hybrid solver for the QUBO model calculations, and excluded the lower performing simulated annealing from the experimental results. When only 10% of the ideal sinogram satisfying the Helgason-Ludwig consistency condition was used, the QCSTR algorithm reconstructed CT images without errors, but the remaining algorithms had difficulty identifying internal structures. Even when random noise was present at around 5% in each pixel in the ideal sinogram, the new algorithm showed good results, as shown in Fig. 8. Since errors are in the form of points within a CT image, it is possible to easily transform it into a shape similar to the sample by removing those points using the connectivity of pixels within the CT image. Of particular note is that tomographic reconstruction to a level where internal identification is possible using only 20% of the data from the 5% noisy sinogram. Due to time limitations in using the hybrid solver, we tested only integers $a$ and $b$ from 1 to 3. We believe that using a more diverse linear combination of two QUBO models will enable tomographic reconstruction with fewer errors. The new algorithm requires different $a$ and $b$ to be combined depending on the type of image, and has not yet found the optimal values for the coefficients. Additionally, the new algorithm has the potential to be further developed as it has the possibility to be combined with additional new QUBO models. When using the QTR algorithm, the most optimized CT image can be obtained when the global minimum energy is satisfied. The problem of not finding the global minimum energy in the ideal sinogram [19] can be solved using Eq. 5 [29]. When using the noisy sinogram, the hybrid solver does not find the global minimum energy, and we are researching a solution to solve this.

QCSTR algorithms can play an important role in medical imaging diagnosis and customized medical device manufacturing. To reconstruct an $nx \times nx$ CT image, typically $nx$ X-ray projection images are required. A new quantum algorithm can achieve error-free tomographic reconstruction with only about 10% of the $nx$ projections. This means that the total amount of radiation a sample receives from a CT scan can be reduced by up to 90%. To examine the internal structure of a $30cm$ sized sample with a pixel size of $0.3mm$, $1000 \times 1000$ CT images must be reconstructed. For optimal tomographic reconstruction, approximately 1000 projection images are required to satisfy the Helgason-Ludwig consistency condition. The new algorithm requires only about 100 projection images to reconstruct the same CT image. When the pixel size required for image diagnosis of the same sample is $1mm$, the new algorithm requires only 30 projection images to reconstruct the CT image. These values can greatly reduce the amount of radiation a patient receives during a CT scan. In particular, the QTR algorithm can compute the QUBO model in at most $O(\log_2 n)$ because it uses the quantum linear system algorithm [18].

Additionally, quantum compressed sensing algorithms can be calculated within a few flops through parallel operations. The new algorithm implies that if quantum computers/annealers become sufficiently advanced that tomographic reconstruction can be reconstructed with a QPU solver, it could provide real-time CT images during surgery. At that time, new algorithms that can reduce the total radiation dose will play an important role. Further research is still needed to determine the coefficients $a$ and $b$ used in the linear combination of the two QUBO models for each sample. Additionally, we are conducting further research to obtain higher resolution CT images by applying new algorithms with a limited number of qubits.

## Data Availability

The Python code used in this paper is in the supplementary material.

Supplementary 1. **Python code for experiments on the QCSTR algorithm.**

More detailed code can be found on the author's GitHub site: https://github.com/ktfriends/Quantum_CT_reconstruction/tree/main/QCSTR

## Competing interests

The key idea developed in this paper were funded by QTomo Inc., which holds ownership of the patent for the algorithm.

## Acknowledgement

D-Wave leap quantum cloud service was supported by the Chungbuk Quantum Research Center at Chungbuk National University. A. R., K. K., and K. J. was supported by the MSIT(Ministry of Science and ICT), Korea, under the ITRC(Information Technology Research Center) support program(IITP-RS-2024-00437284) supervised by the IITP(Institute for Information & Communications Technology Planning & Evaluation. A. R. and K. K. was supported by Creation of the quantum information science R&D ecosystem(based on human resources) (Agreement Number) through the National Research Foundation of Korea(NRF) funded by the Korean government (Ministry of Science and ICT(MSIT)) (RS-2023-00256050).

## Supplementary Figures

Supplementary Figure 1. **Density graph**. The upper images are data from the synchrotron radiation accelerator, and the lower images are data to which random noise with a Gaussian distribution was applied for this experiment.

Supplementary Figure 2. **Difference between each sample and CT image in Figure 8.**